\documentclass[11pt,a4paper]{article}
\usepackage[utf8]{inputenc}
\usepackage{amsmath}
\usepackage{amsfonts}
\usepackage{hyperref}
\usepackage{amssymb}
\usepackage{makeidx}
\usepackage{cite}
\usepackage{bm}
\usepackage{geometry}
\usepackage{doc}

\usepackage{url}

\usepackage{graphicx}
\usepackage{epstopdf}
\begin{document}
\begin{titlepage}

~~\\

\vspace*{0cm}
    \begin{Large}
    \begin{bf}
       \begin{center}
         {Modular Hamiltonians and large diffeomorphisms in $AdS_{3}$}
       \end{center}
    \end{bf}   
    \end{Large}
    
  \vspace{0.7cm}
\begin{center}   
Suchetan Das\footnote
            {
e-mail address : 
suchetan.das@rkmvu.ac.in},
Bobby Ezhuthachan\footnote
            {
e-mail address : 
bobby.ezhuthachan@rkmvu.ac.in}

\vspace{0.3cm}
 {\it Ramakrishna Mission Vivekananda Educational and Research Institute, Belur Math, Howrah-711202, West Bengal, India}
\end{center}

\vspace{0.7cm}
\begin{abstract}
We generalize the work of Kabat and Lifshytz (arXiv:1703.06523), of reconstructing bulk scalar fields using the intersecting modular Hamiltonian approach discussed therein, to any locally $AdS_3$ space related to $AdS_3$ by large diffeomorphisms. We present several checks for our result including matching with their result in appropriate limits as well as consistency with bulk diffeomorphisms. As a further check, from our expressions we also compute the first correction due to gravitational dressing to the bulk scalar field in $AdS_3$ and match with known results in the literature.
\end{abstract}
 \end{titlepage}

\pagenumbering{arabic}

\tableofcontents
\section{Introduction}
The $AdS$/CFT correspondence \cite{Maldacena:1997re}-\cite{Aharony:1999ti} implies an isomorphism of Hilbert spaces of a Quantum Gravity theory ($\mathcal{H}_{B}$) on asymptotically $AdS$ spacetime(A$Ads$) and a QFT ($\mathcal{H}_b$) on its  conformal boundary. In particular, the bulk and boundary states are identified. In terms of the density matrices or its logarithm($K$), this would mean $K_{B} = K_b$ \footnote{ where the subscripts $B$ and $b$ denote the bulk and boundary respectively.}. 
When the Hilbert space is factorizable into  $\mathcal{H}_{b}=\mathcal{H}^{a}_b\otimes \mathcal{H}^{a^c}_b$, the above identification of states would naively suggest an equality of reduced density matrices ($K^{a}_{b} =K^{a'}_B$), where $\mathcal{H}^{a'}_{B}$ is the image under the isomorphic map from $\mathcal{H}^{a}_{b}$.  Since the map is not known, except in the semi-classical limit, in general one does not know what $\mathcal{H}^{a'}_{B}$ is for a given $\mathcal{H}^a_{b}$. In the semi classical limit though, when the bulk states can be approximated by a spacetime geometry, a concrete identification was proposed in \cite{Jafferis:2015del}, building on the previous important works of[\cite{Ryu:2006bv}, \cite{Lewkowycz:2013nqa}, \cite{Faulkner:2013ana}]

\begin{equation}\label{JLMS}
K^a_{b} = K^{a'}_{B} + \hat{A}_{ext}/G_N + \hat{S}_{wald}/G_{N} + ... + \mathcal{O}(G_N) 
\end{equation}

In a local QFT, the factorization of the boundary Hilbert space is assumed to happen across a spatial region ie `$a$' denotes a spatial region, with a boundary $\partial a$\footnote{This is not strictly correct and the right way to describe entanglement in QFT is in terms of algebra of observables. See the nice review \cite{Witten:2018zxz} and watch\cite{youtube} for a nice discussion of this point.}. In this semi-classical limit, the corresponding bulk region `$a'$' is the spatial region bounded by $\partial a$ and the Ryu-Takayanagi minimal surface ending on $\partial a$. It is interesting to note, that RHS in (\ref{JLMS}) has contributions from operators localized on the RT surface \cite{Ryu:2006bv},\cite{Hubeny:2007xt}, including the Area operator, which infact dominates in $G_N\rightarrow 0$ limit\footnote{See \cite{Harlow:2016vwg}, \cite{Harlow:2018fse} for a nice discussion on the appearance of the Area term in the RHS of (\ref{JLMS}) and a derivation of the result in the language of quantum error correction\cite{Almheiri:2014lwa}. }.   

Equation (\ref{JLMS}) has implications for the bulk reconstruction of local fields as smeared operators in the boundary. This was pointed out in\cite{Jafferis:2015del} itself and discussed in detail in\cite{Faulkner:2017vdd}, wherein it was showed, how the above identification can be used to reconstruct bulk operators inside the entanglement wedge- ($\mathcal{D}(a')$) from operators localized inside $a$\footnote{See also \cite{Dong:2016eik} for a proof of the entanglement wedge reconstruction in the language of quantum error correcting codes and also \cite{Sanches:2017xhn} for related work.}. Independently, it was shown in \cite{Kabat:2017mun} that, the above equation can be used to reproduce the HKLL smeared representation of local bulk fields [\cite{Hamilton:2005ju}-\cite{Hamilton:2006fh}] in the causal wedge for $AdS_3$  and the BTZ metric. 

In \cite{Faulkner:2017vdd} it was shown that a bulk local field inside the entanglement wedge ($\mathcal{D}(a')$) can be expressed as a smeared integral over the Fourier modular transforms (Fourier transform of modular flows) of boundary operators ($\mathcal{O}$), localised over the spatial region $(a)$\footnote{see\cite{Chen:2018rgz} and \cite{Faulkner:2018faa} for recent applications of modular flows to the bulk reconstruction program.}. The Fourier modular transform is defined as:
\begin{equation}
\mathcal{O}_{\omega} = \int ds e^{i\omega s} \mathcal{O}_{s}, \; \;  \mathcal{O}_{s} = e^{isK_b}\mathcal{O} e^{-isK_b}
\end{equation}
They are by definition eigenmodes of the modular Hamiltonian $K_b$ 
\begin{equation}
[K_b, \mathcal{O}_{\omega}] = \omega \mathcal{O}_{\omega}
\end{equation}
In particular, the zero mode $\mathcal{O}_{0}$ commutes with the modular Hamiltonian\footnote{\cite{Faulkner:2017vdd} also provides a formula for the zero modes as integral of the bulk field over the RT surface, thus generalizing the OPE block formula of\cite{Czech:2016xec} beyond the vacuum.}.

In an interesting paper \cite{Kabat:2017mun}, Kabat and Lifshytz were able to reproduce the HKLL formula for the free local bulk field in $AdS_3$ and BTZ metric, using equation (\ref{JLMS}).  The basic idea is to use the fact that, a bulk scalar field localized on the RT surface, would commute with the boundary Modular Hamiltonian, as an equation to solve for the bulk field \footnote{In\cite{Kabat:2017mun}, this condition was argued to follow from the identification of the bulk and boundary modular Hamiltonians as well as the fact that the bulk RT surface is invariant under the action of the bulk modular Hamiltonian. This condition also follows from the analysis on zero-modes presented in section-II of \cite{Faulkner:2017vdd}. In particular, it follows from equation (4.48) of that paper.}. Taking two intersecting bulk geodesics and imposing this condition on the bulk field localised at the intersection point of the geodesics and simultaneously solving both the conditions provides an explicit form for the bulk field\footnote{Also this procedure could help to extract the form of bulk metric within the causal wedge as described in \cite{Roy:2018ehv}.}. In this note we generalize their method for a bulk field in any locally $AdS_3$ metric, which are related to $AdS_3$ by large diffeomorphisms. 

We present the details of the analysis in the next section. The smeared integral representation for the bulk field is given in equation (\ref{result}). In section (\ref{checks}) we provide several checks of our calculation.  We show in section (\ref{splcase}), that our result reduces to the expressions for the bulk field in $AdS_3$ and the BTZ geometry in the right limit. We also show in section (\ref{diffeo check}) and section (\ref{bulkboundary}), that the result we obtain is consistent with bulk diffeomorphism using some of the methods developed in \cite{Anand:2017dav}. Finally in section (\ref{TO correction}), we look at small perturbations from the $AdS_3$ metric. In this limit we show that the expression for the bulk field can be reorganized as a correction over the $AdS_3$ field. The correction term incorporates the gravitational dressing due to interaction with the stress-energy tensor. We explicitly compute the first correction due to the stress tensor interaction and match it with known results in the literature \cite{Guica:2016pid}\footnote{See also \cite{Lewkowycz:2016ukf}.}.

\section{Bulk field from intersecting modular Hamiltonians}\label{modham}
In this section, we generalize the discussion in \cite{Kabat:2017mun} to any locally $AdS_3$ metric in the bulk. A large diffeomorphism in the bulk would correspond to a general conformal transformation on the boundary. We are therefore interested in the Modular Hamiltonian for the excited state, obtained by local conformal transformations on the vacuum.

Now the Modular Hamiltonian of a single interval on a constant time slice, in a 2D CFT \cite{Hislop:1981uh}-\cite{Cardy:2016fqc}, is given by 

\begin{align}\label{modvac}
H = \int_{U}^{V} d\omega \frac{(\omega - U)(V-\omega)}{V-U} T_{\omega\omega}(\omega) + \int_{U}^{V} d\bar{\omega} \frac{(\bar{\omega} - U)(V-\bar{\omega})}{V-U} \bar{T}_{\bar{\omega}\bar{\omega}}(\bar{\omega})
\end{align}

To find the expression for the other locally excited state in the CFT, we need to perform a conformal transformation $(\omega,\bar{\omega}) \rightarrow g(\omega),\bar{g}(\bar{\omega})$\footnote{Since, originally the two points of the interval are at constant time($U=\bar{U}$), they will remain unchanged after such boundary diffeomorphism($U,\bar{U}\rightarrow g(u),\bar{g}(\bar{u})$), i.e $\bar{g}(\bar{u}) = g(u)$. But $u$ and $\bar{u}$ are not necessarily same.} 
The action of this map on (\ref{modvac}) is straightforward. Under this map, the stress tensor is transformed as,
\begin{align}
T(g(\omega))=\frac{T(\omega)}{g'^{2}(\omega)}-\frac{c}{12}\frac{\mathcal{S}\{g(\omega),\omega\}}{g'^{2}(\omega)}
\end{align}
Where $\mathcal{S}\{g(\omega),\omega\} \equiv \frac{g'''}{g'} -\frac{3}{2}\frac{g''^{2}}{g'^{2}}$ is the Schwarzian derivative. This Schwarzian part of the integral gives some constant which we can ignore since we are ultimately interested in the commutator of modular Hamiltonian with local operator. The non trivial part, which contributes in the commutator is involving the stress tensor part. In this part, $T(g(\omega))$ is related to $T(\omega)$ through a multiplication factor of $\frac{1}{g^{'2}(\omega)}$ and get cancelled to the factor coming from the Jacobian $g'(z)$, which in turn reduces to the following expression.

\begin{align}\label{modexc}
\tilde{H}_{mod} = \tilde{H}_{mod}^{(R)} + \tilde{H}_{mod}^{(L)} &= \int_{-\infty}^{+\infty} d\omega \frac{(g(\omega) - g(u))(g(v)-g(\omega))}{g'(\omega)(g(v)-g(u))} T_{\omega\omega}(\omega) + \nonumber \\
& + \int_{-\infty}^{+\infty} d\bar{\omega} \frac{(\bar{g}(\bar{\omega}) - g(u))(g(v)-\bar{g}(\bar{\omega}))}{\bar{g}'(\bar{\omega})(g(v)-g(u))} \bar{T}_{\bar{\omega}\bar{\omega}}(\bar{\omega})
\end{align}
Here, instead of $H_{mod}$, we define total modular Hamiltonian $\tilde{H}_{mod}$ by extending the limit of the of the integration where it is non zero even outside of the segment.\footnote{More technically, the total modular Hamiltonian is defined by the difference of modular Hamiltonian of a subsystem $A$ and that of it's compliment subsystem $A^{c}$. $\tilde{H}_{mod} = H_{mod}^{A}-H_{mod}^{A^{c}}$. In stead of putting $[H_{A},\phi]=0$, we impose $[\tilde{H},\phi]=0$, as originally done in \cite{Kabat:2017mun}. This will lead to an expression for $\phi$ with a support from the full $AdS$ spacetime(unlike the entanglement wedge reconstruction). It also manifests the fact that RT surface acts as a bifurcation surface between $A$ and $A^{c}$.}\\
In the Lorentzian CFT$_{2}$ we have the following commutators of stress tensor and primary scalar operator $\mathcal{O}$ with $h=\bar{h}=\frac{\Delta}{2}$.
\begin{align}
2\pi[T(\omega),\mathcal{O}(\zeta,\bar{\zeta})] = 2\pi i(h\partial_{\zeta}\delta(\zeta-\omega)+\delta(\zeta-\omega)\partial_{\zeta})\mathcal{O}(\zeta,\bar{\zeta}), \nonumber \\
2\pi[\bar{T}(\bar{\omega}),\mathcal{O}(\zeta,\bar{\zeta})] = -2\pi i(h\partial_{\bar{\zeta}}\delta(\bar{\zeta}-\bar{\omega})+\delta(\bar{\zeta}-\bar{\omega}) \partial_{\bar{\zeta}})\mathcal{O}(\zeta,\bar{\zeta})
\end{align}
Using these commutators we can easily find the commutators between modular Hamiltonian and the primary scalar operator.
\begin{align}
[\tilde{H}_{mod}^{(R)},\mathcal{O}(\zeta,\bar{\zeta})] &= \frac{2\pi i}{(g(v)-g(u))} [ h(g(v)+g(u)-2g(\zeta)) - \nonumber \\
& -(g(\zeta)-g(u))(g(v)-g(\zeta))\left(\frac{hg''(\zeta)}{g'^{2}(\zeta)} - \frac{\partial_{\zeta}}{g'(\zeta)}\right)]\mathcal{O}(\zeta,\bar{\zeta})
\end{align}
\begin{align}
[\tilde{H}_{mod}^{(L)},\mathcal{O}(\zeta,\bar{\zeta})] & = -\frac{2\pi i}{(g(v)-g(u))}[h(g(v)+g(u)-2\bar{g}(\bar{\zeta})) - \nonumber \\
&-(\bar{g}(\bar{\zeta})-g(u))(g(v)-\bar{g}(\bar{\zeta}))\left(\frac{h\bar{g}''(\bar{\zeta})}{\bar{g}'^{2}(\bar{\zeta})} - \frac{\partial_{\bar{\zeta}}}{\bar{g}'(\bar{\zeta})}\right)]\mathcal{O}(\zeta,\bar{\zeta})
\end{align}
Therefore the commutator of total total modular Hamiltonian and the primary becomes,
\begin{align}
[\tilde{H}_{mod},\mathcal{O}(\zeta,\bar{\zeta})] & =\frac{2\pi i}{(g(v)-g(u))} [\Delta(\bar{g}(\bar{\zeta})-g(\zeta)) + g(u)g(v)(h\frac{g''(\zeta)}{g'^{2}(\zeta)}-h\frac{\bar{g}''(\bar{\zeta})}{\bar{g}'^{2}(\bar{\zeta})} - \frac{\partial_{\zeta}}{g'(\zeta)}+\frac{\partial_{\bar{\zeta}}}{\bar{g}'(\zeta)}) \nonumber \\
& -(g(u)+g(v))(h\frac{g(\zeta)g''(\zeta)}{g'^{2}(\zeta)}-h\frac{\bar{g}(\bar{\zeta})\bar{g}''(\bar{\zeta})}{\bar{g}'^{2}(\bar{\zeta})} - \frac{g(\zeta)\partial_{\zeta}}{g'(\zeta)}+\frac{\bar{g}(\bar{\zeta})\partial_{\bar{\zeta}}}{\bar{g}'(\zeta)}) + \nonumber \\
& + (h\frac{g^{2}(\zeta)g''(\zeta)}{g'^{2}(\zeta)}-h\frac{\bar{g}^{2}(\bar{\zeta})\bar{g}''(\bar{\zeta})}{\bar{g}'^{2}(\bar{\zeta})} - \frac{g^{2}(\zeta)\partial_{\zeta}}{g'(\zeta)}+\frac{\bar{g}^{2}(\bar{\zeta})\partial_{\bar{\zeta}}}{\bar{g}'(\zeta)})] \mathcal{O}(\zeta,\bar{\zeta})
\end{align}
\subsection{Smearing function ansatz for the bulk field}\label{smearing}
Following\cite{Kabat:2017mun}, we would like to find an integral expression for bulk scalar field. Our ansatz for the bulk field at some bulk point $X$(which is, as of now, a free CFT variable) is
\begin{align}
\phi(X) = \int dp dq K(p,q)\mathcal{O}(p,q)
\end{align}
where we choose the following parametrization, $q=\zeta +iy'$ and $p=\bar{\zeta}+iy'$. $\zeta$ and $\bar{\zeta}$ are CFT variables in lightcone coordinates and $y'$ is also some parametrization of boundary CFT variable in some particular conformal frame. $K(p,q)$ is the kernel which we would like to derive for this case. We will comment on this parametrization at the end of this section and in the section (\ref{splcase}), we will see some choices of such parametrization in two particular examples. The condition from CFT is that the total modular Hamiltonian commutes with the bulk scalar field on the RT surface(in this case which is a geodesic connecting $(u,v)$). i.e
\begin{align}\label{commut}
&[\tilde{H}^{(u,v)}_{mod},\phi] = \int dp dq K(p,q)[\tilde{H}^{(u,v)}_{mod},\mathcal{O}(p,q)] = 0 \\ 
&\implies [(\Delta-2)(\bar{g}(p)-g(q)) + (h-1)[g^{2}(q)\frac{g''(q)}{g'^{2}(q)}-\bar{g}^{2}(p)\frac{\bar{g}''(p)}{\bar{g}'^{2}(p)} + \nonumber \\
& g(u)g(v)(\frac{g''(q)}{g'^{2}(q)}-\frac{\bar{g}''(p)}{\bar{g}'^{2}(p)}) + 
- (g(u)+g(v))(g(q)\frac{g''(q)}{g'^{2}(q)}-\bar{g}(p)\frac{\bar{g}''(p)}{\bar{g}'^{2}(p)})]]K(p,q)  \nonumber \\
& = [ (g(u)+g(v))(\frac{g(q)\partial_{q}}{g'(q)} - \frac{\bar{g}(p)\partial_{p}}{\bar{g}'(p)}) + 
-g(u)g(v)(\frac{\partial_{q}}{g'(q)} - \frac{\partial_{p}}{\bar{g}'(p)}) - (\frac{g^{2}(q)\partial_{q}}{g'(q)} - \frac{\bar{g}^{2}(p)\partial_{p}}{\bar{g}'(p)})]K(p,q)
\end{align}
But to specify the bulk field localized in a bulk point, we need to consider two intersecting geodesics and in the intersection point the field is localized. Therefore the modular Hamiltonian must commute with the bulk operator localized on the two geodesics with endpoints $(y_{1},y_{2}),(y_{3},y_{4})$. The difference between this two condition gives:
\begin{align}
(g(y_{1})-g(y_{2}))[\tilde{H}^{(y_{1},y_{2})}_{mod},\phi] - (g(y_{3})-g(y_{4}))[\tilde{H}^{(y_{3},y_{4})}_{mod},\phi] = 0 
\end{align}
Doing some simple algebra and integrating by-parts we get the following equation
\begin{align}
\left[\left(\frac{(X_{0}-g(q))}{g'(q)}\partial_{q} - \frac{(X_{0}-\bar{g}(p))}{\bar{g}'(p)}\partial_{p}\right) - (h-1)\left(\frac{(g(q)-X_{0})g''(q)}{g'^{2}(q)} - \frac{(\bar{g}(p)-X_{0})\bar{g}''(p)}{\bar{g}'^{2}(p)}\right)\right]K(p,q) = 0
\end{align}
where $X_{0}=\frac{g(y_{1})g(y_{2})-g(y_{3})g(y_{4})}{g(y_{1})+g(y_{2})-g(y_{3})-g(y_{4}}$. Using the method of characteristics we have,
\begin{align}
\frac{g'(q)dq}{(X_{0}-g(q))} = -\frac{\bar{g}'(p)dp}{(X_{0}-\bar{g}(p))} = \frac{dK(p,q)}{(h-1)\left(\frac{(g(q)-X_{0})g''(q)}{g'^{2}(q)} - \frac{(\bar{g}(p)-X_{0})\bar{g}''(p)}{\bar{g}'^{2}(p)}\right)K(p,q)}
\end{align}
The first two equation gives the equation of the characteristic curve, which is
\begin{align}
(g(q)-X_{0})(\bar{g}(p)-X_{0}) = A
\end{align}
for some constant $A$. This curve equation implies
\begin{align}
\frac{g''(q)}{g'^{2}(q)} = \frac{1}{g(q)-X_{0}}, \quad \frac{\bar{g}''(p)}{\bar{g}'^{2}(p)} = \frac{1}{\bar{g}(p)-X_{0}}
\end{align}
Along this characteristic curve we have to solve the rest of the equation to find the general solution for $K(p,q)$,
\begin{align}
\int \frac{dK}{K} = -(h-1)\int \frac{g''(q)dq}{g'(q)} -(h-1)\int \frac{g'(q)}{g(q)-X_{0}} \nonumber \\
\implies \ln K = -\ln g'(q)^{h-1} -\ln (g(q)-X_{0})^{h-1} + c_{1}(p,A)
\end{align}
similarly from the other part we also get,
\begin{align}
 \ln K = -\ln \bar{g}'(p)^{h-1} -\ln (\bar{g}(p)-X_{0})^{h-1} + c_{2}(q,A)
\end{align}
Therefore we have the most general solution of $K(p,q)$ is
\begin{align}
K(p,q) = (g'(q)\bar{g}'(p))^{1-h}H((g(q)-X_{0})(\bar{g}(p)-X_{0}))
\end{align}
Where $H(z)$ is some arbitrary function of $z$ and $z=(g(q)-X_{0})(\bar{g}(p)-X_{0})$.
Putting this solution to (\ref{commut}) we end up with the following equation
\begin{align}
[z-g(u)g(v) + (g(u)+g(v))X_{0} -X_{0}^{2}]\frac{dH(z)}{dz} = (\Delta-2)H(z)
\end{align}
One can solve it simply and finally get the following expression for the kernel $K(p,q)$.
\begin{align}
K(p,q) = c_{\Delta} \left( \frac{(g(q)-X_{0})(\bar{g}(p)-X_{0})+\tilde{Y}^{2}}{\sqrt{g'(q)\bar{g}'(p)}}\right)^{\Delta-2}
\end{align}
where $\tilde{Y}^{2} =(g(u)+g(v))X_{0} -X_{0}^{2}-g(u)g(v)$, is the bulk radial coordinate(we will justify this in the next section). To do the integration by parts without any boundary terms, we need the integration region to be bounded,$(g(q)-X_{0})(\bar{g}(p)-X_{0})+\tilde{Y}^{2} > 0$. Thus, as expected, we have fixed the local bulk operator upto a spacetime dependent coefficient $c_{\Delta}$. For the states where CFT has spacetime translation symmetry, the coefficient is a function of bulk radial coordinate $\tilde{Y}$ only. Using the $AdS_{3}$/CFT$_{2}$ boundary condition $\phi((\tilde{Y}\rightarrow 0),X) \rightarrow \frac{\tilde{Y}^{\Delta}}{2\Delta-2}\mathcal{O}(X)$, we will fix $c_{\Delta}$ and get,
\begin{align}\label{result}
\phi(X,Y)=\int_{(g(q)-X_{0})(\bar{g}(p)-X_{0})+\tilde{Y}^{2} > 0}dq dp \left( \frac{(g(q)-X_{0})(\bar{g}(p)-X_{0})+\tilde{Y}^{2}}{\tilde{Y}\sqrt{g'(q)\bar{g}'(p)}}\right)^{\Delta-2} \mathcal{O}(q,p)
\end{align}
One can further reparametrize $q$ and $p$ in terms of CFT variables in some conformal frame. In the next subsection, we will see how to do that in some specific cases. Nevertheless, the correct choice of parametrization of $q,p$ comes from demanding it to be real such that, $(g(q)-X_{0})(\bar{g}(p)-X_{0})+\tilde{Y}^{2}$ is bounded above zero. We will also see in the section (\ref{diffeo check}) that, $X_{0}$ is indeed the bulk point $X$ where the two geodesics intersect. 
\section{Checks on the calculation}\label{checks}

In this section, as a check of our calculation above, we show that the expression for the bulk local field given in (\ref{result}), reduces to the known results for $AdS_3$ and the BTZ cases. We also show that it is consistent with what one expects from bulk large diffeomorphisms. 
\subsection{Reproducing the expression for  $AdS_3$ and BTZ}\label{splcase}

As a first check of our calculation in the previous section, We now show how the above result reduces to the known results in the pure $AdS_3$ and BTZ cases.
\begin{itemize}
\item Free scalar field in Pure vacuum $AdS$ :
\end{itemize}
It is the simplest case, where we can choose the parametrization of $q$ and $p$ as follows, $g(q)=q=x-t'+iy',\bar{g}(p)=p=x+t'+iy'$. It reduces the smearing Kernel $K(p,q)$ to,
\begin{align}
K(p,q) = \left(\frac{y^{2}+(q-X_{0})(p-X_{0})}{y}\right)^{\Delta-2}
\end{align}
Here $y^{2} = (u+v)X_{0}-uv+X_{0}^{2}$. It is the equal-time geodesic equation for pure $AdS$ vacuum in Poincare coordinates$(y,z,\bar{z})$. Hence we can recover the famous HKLL formula \cite{Hamilton:2005ju}-\cite{Hamilton:2006fh}:
\begin{align}
\phi(y,x,T=0) = \frac{\Delta-1}{\pi} \int_{y'^{2}+t'^{2}<y^{2}} \left(\frac{y^{2}-y'{^2}-t'^{2}}{y}\right)^{\Delta-2}\mathcal{O}(t',x+iy')
\end{align}
\begin{itemize}
\item Free scalar field in BTZ background:
\end{itemize}
In this case $g(q) = e^{\frac{r_{+}q}{l^{2}}} = e^{\tilde{q}}$ and $\bar{g}(p)=e^{\tilde{p}}$. We chose $(g(y_{1}),g(y_{2}))=(e^{\tilde{R}},e^{-\tilde{R}})$ and $(g(y_{3}),g(y_{4}))=(e^{\tilde{\phi_{0}}+\tilde{L}},e^{\tilde{\phi_{0}}-\tilde{L}})$. Thus
\begin{align}
X_{0} &= \frac{1-e^{2\tilde{\phi_{0}}}}{e^{\tilde{R}}+e^{-\tilde{R}}-e^{\tilde{\phi_{0}}+\tilde{L}}-e^{\tilde{\phi_{0}}-\tilde{L}}} \nonumber \\
& = -\frac{1}{\frac{\cosh\tilde{R}}{\sinh\tilde{\phi}_{0}}\left(\cosh\tilde{\phi}_{0}-\frac{\cosh\tilde{L}}{\cosh\tilde{R}}-\sinh\tilde{\phi}_{0}\right)} \nonumber \\
& = -\frac{1}{\cosh\tilde{R}\tanh\tilde{\phi_{*}}-\cosh\tilde{R}} \nonumber \\
& = \frac{1+e^{2\tilde{\phi_{*}}}}{2\cosh\tilde{R}}
\end{align}
Where we define $\tanh\tilde{\phi_{*}} = \frac{1}{\sinh\tilde{\phi}_{0}}\left(\cosh\tilde{\phi}_{0}-\frac{\cosh\tilde{L}}{\cosh\tilde{R}}\right)$. Now the smearing function is,
\begin{align}
K(p,q) = \left[\frac{e^{\frac{\tilde{q}+\tilde{p}}{2}}+(X_{0}^{2} +\tilde{Y}^{2})e^{-\frac{\tilde{q}+\tilde{p}}{2}} -X_{0}\left(e^{\frac{\tilde{q}-\tilde{p}}{2}}+e^{\frac{\tilde{p}-\tilde{q}}{2}}\right)}{\tilde{Y}}\right]^{\Delta-2}
\end{align}
Here $\tilde{Y}^{2}+X_{0}^{2} = (g(u)+g(v))X_{0}-g(u)g(v) = e^{2\tilde{\phi_{*}}}$. So $\tilde{Y} = e^{\tilde{\phi_{*}}}\sqrt{\left(1-\frac{\cosh^{2}\tilde{\phi_{*}}}{\cosh^{2}\tilde{R}}\right)} $Therefore,
\begin{align}
K(p,q) &= \left[\frac{e^{\frac{\tilde{q}+\tilde{p}}{2}}+e^{2\tilde{\phi_{*}}-\frac{\tilde{q}+\tilde{p}}{2}}-\left(\frac{1+e^{2\tilde{\phi_{*}}}}{2\cosh\tilde{R}}\right)\cosh\left(\frac{\tilde{p}-\tilde{q}}{2}\right)}{\tilde{Y}}\right]^{\Delta-2} \nonumber \\
& = \left[\frac{\cosh\left(\frac{\tilde{q}+\tilde{p}}{2}-\tilde{\phi_{*}}\right) - \frac{\cosh\tilde{\phi_{*}}}{\cosh\tilde{R}}\cosh\left(\frac{\tilde{p}-\tilde{q}}{2}\right)}{\tilde{Y}e^{-\tilde{\phi_{*}}}}\right]^{\Delta-2}
\end{align}
On a constant time slice, the geodesic of BTZ blackhole \footnote{Here, the metric of BTZ is,
\begin{align}
ds^{2} = -\frac{r^{2}-r_{+}^{2}}{l^{2}}dt^{2}+\frac{l^{2}}{r^{2}-r_{+}^{2}}dr^{2}+r^{2}d\phi^{2}
\end{align}
}, connecting the two boundary points at $(-\tilde{R},\tilde{R})$, satisfies the following relation
\begin{align}
\sqrt{1-\frac{r_{+}^{2}}{r^{2}}} = \frac{\cosh\tilde{\phi_{*}}}{\cosh\tilde{R}}
\end{align}
Where $r$ is the bulk radial coordinate. Therefore, the bulk coordinate can be written in terms of radial coordinate $r$. $\tilde{Y} = \frac{r_{+}}{r} e^{\tilde{\phi_{*}}}$. Now we replace $p,q$ with the following parametrization in terms of boundary spacetime variables of BTZ metric, such that
\begin{align}
\tilde{q} = \frac{r_{+}}{l^{2}}(l\phi - \frac{l^{2}x}{r_{+}} + i\frac{l^{2}y}{r_{+}}), \quad \tilde{p} = \frac{r_{+}}{l^{2}}(l\phi + \frac{l^{2}x}{r_{+}} + i\frac{l^{2}y}{r_{+}})
\end{align}
 In this way we recover the well-known BTZ bulk field in terms of boundary CFT operator \cite{Kabat:2017mun}.
 \begin{align}
 \phi^{(0)}(r,\phi,t=0) \sim \left(\frac{r}{r_{+}}\right)^{\Delta-2} \int dx dy \left(\cos y - \sqrt{1-\frac{r_{+}^{2}}{r^{2}}}\cosh x \right)^{\Delta-2} \mathcal{O}\left(\phi + \frac{ily}{r_{+}}, \frac{l^{2}x}{r_{+}}\right)
 \end{align}

\subsection{Consistency with bulk diffeomorphisms}\label{diffeo check}
In this section, we want to connect this result with that obtained from bulk diffeomorphism. Let us consider the general form of large diffeomorphism from vacuum $AdS_{3}$ uniformizing coordinates $(y,z,\bar{z})$ to arbitrary $(Y,Z,\bar{Z})$ via:
\begin{align}\label{diff}
z =& g(Z) + G(g(Z),\bar{g}(\bar{Z}),Y) \quad \ni \quad G(g(Z)=Z,\bar{g}(\bar{Z}),Y) = 0; \quad G(g(Z),\bar{g}(\bar{Z}),Y=0) = 0  \nonumber \\
\bar{z} =& \bar{g}(\bar{Z}) + \bar{G}(g(Z),\bar{g}(\bar{Z}),Y) \quad \ni \quad \bar{G}(g(Z),\bar{g}(\bar{Z})=\bar{Z},Y) = 0; \quad G(g(Z),\bar{g}(\bar{Z}),Y=0) = 0  \nonumber \\
y =& YF(g(Z),\bar{g}(\bar{Z}),Y) = \tilde{Y} \quad \ni \quad F(g(Z)=Z,\bar{g}(\bar{Z})=\bar{Z},Y) = 1
\end{align}
Here, $G,\bar{G}$ are the arbitrary functions of boundary Virasoro transformation $g,\bar{g}$ and bulk coordinate $Y$. This general form suggests that, as $z=\bar{z}$ at $y=0$, $g(Z)=\bar{g}(\bar{Z})|_{Y=0}$. Therefore, at the boundary of the new metric, the Virasoro transformation acts in a similar way to that of the $AdS$. First we see how the geodesic in $AdS$ at constant time changes under diffeomorphism. For two boundary points $(y_{1},y_{2})$ at $T=0$ slice, the geodesic is,
\begin{align}
y^{2} = (y_{1}+y_{2})\frac{z+\bar{z}}{2}|_{z=\bar{z}} - y_{1}y_{2} - (\frac{z+\bar{z}}{2})^{2}|_{z=\bar{z}}
\end{align}
Using the general form of diffeomorphism (\ref{diff}), we get,
\begin{align}\label{12}
\tilde{Y}^{2} &= (g(Y_{1})+g(Y_{2}))(g(Z)+G) - g(Y_{1})g(Y_{2}) - (g(Z)+G)^{2} \\
& = (g(Y_{1})+g(Y_{2}))(\bar{g}(\bar{Z})+\bar{G}) - g(Y_{1})g(Y_{2}) - (\bar{g}(\bar{Z})+\bar{G})^{2} \nonumber
\end{align}
similarly for boundary points $(y_{3},y_{4})$ in uniformizing coordinate, we also have
\begin{align}\label{34}
\tilde{Y}^{2} &= (g(Y_{3})+g(Y_{4}))(g(Z)+G) - g(Y_{3})g(Y_{4}) - (g(Z)+G)^{2} \\
& = (g(Y_{3})+g(Y_{4}))(\bar{g}(\bar{Z})+\bar{G}) - g(Y_{3})g(Y_{4}) - (\bar{g}(\bar{Z})+\bar{G})^{2} \nonumber
\end{align}
Where under Virasoro transformation $(y_{1},y_{2},y_{3},y_{4}) \rightarrow (Y_{1},Y_{2},Y_{3},Y_{4})$. Taking the difference of (\ref{12}) and (\ref{34}), we can get the point at which these two geodesics intersect. That is,
\begin{align}\label{intersection}
g(Z)+G=\bar{g}(\bar{Z})+\bar{G} = \frac{g(Y_{1})g(Y_{2})- g(Y_{3})g(Y_{4})}{g(Y_{1})+g(Y_{2})-g(Y_{3})-g(Y_{4})} = X_{0}
\end{align}
At this point of intersection, the geodesic (\ref{12}) is of the form, as we got in the previous section
\begin{align}
\tilde{Y}^{2} = (g(Y_{1})+g(Y_{2}))X_{0}- g(Y_{1})g(Y_{2}) - X_{0}^{2} 
\end{align}
Let us now look how the local bulk operator in the uniformizing coordinate will be modified under such diffeomorphism. The $AdS_{3}$ free field is 
\begin{align}
\phi(y,z,\bar{z}) = \int dq dp \left(\frac{(q-z)(p-\bar{z})+y^{2}}{y}\right)^{\Delta-2}\mathcal{O}(q,p)
\end{align}
Using (\ref{diff}) we have,
\begin{align}
\phi(\tilde{Y},Z,\bar{Z}) &= \int dg(Q) d\bar{g}(P) \left(\frac{(g(Q)-(g(Z)+G))(\bar{g}(P)-(\bar{g}(\bar{z})+\bar{G}))+\tilde{Y}^{2}}{\tilde{Y}}\right)^{\Delta-2}\mathcal{O}(g(Q),\bar{g}(p)) \nonumber \\
& = \int dQ dP \left(\frac{(g(Q)-X_{0})(\bar{g}(P)-X_{0})+\tilde{Y}^{2}}{\tilde{Y}\sqrt{g'(Q)\bar{g}'(P)}}\right)^{\Delta-2}\mathcal{O}(Q,P)
\end{align}
Here in the last line we used (\ref{intersection}) and the property of scalar primary operator $\mathcal{O}(g(Q),\bar{g}(P))$. Hence we reproduced the same result, obtained in (\ref{result}).

\subsection{Bulk-boundary OPE block and propagator}\label{bulkboundary}
In this section we calculate the vacuum sector OPE block, constructed out of the OPE of one bulk scalar field and one boundary scalar primary operator and  match it with the expected result from diffeomorphism of $AdS_{3}$. For simplicity, we consider the bulk field located at the center($Z=\bar{Z}=0$). Hence,

\hspace*{-0.6cm}\vbox{\begin{align}
&\phi(\tilde{Y},0,0)\mathcal{O}(\zeta,\bar{\zeta})|_{vac} \nonumber \\
& = \int_{(g(q)-X_{0})(\bar{g}(p)-X_{0})+\tilde{Y}^{2}>0} dq dp \left(\frac{(g(q)-X_{0})(\bar{g}(p)-X_{0})+\tilde{Y}^{2}}{\tilde{Y}\sqrt{g'(q)\bar{g}'(p)}}\right)^{\Delta-2}\mathcal{O}(q,p)\mathcal{O}(\zeta,\bar{\zeta})|_{vac} \nonumber \\
& = \int dg(q) d\bar{g}(p)(g'(\zeta)\bar{g}'(\bar{\zeta}))^{\frac{\Delta}{2}} \left(\frac{(g(q)-X_{0})(\bar{g}(p)-X_{0})+\tilde{Y}^{2}}{\tilde{Y}}\right)^{\Delta-2}\mathcal{O}(g(q),\bar{g}(p))\mathcal{O}(g(\zeta),\bar{g}(\bar{\zeta}))|_{vac} \nonumber \\
& = (g'(\zeta)\bar{g}'(\bar{\zeta}))^{\frac{\Delta}{2}} \int_{\tilde{g}\tilde{\bar{g}}+\tilde{Y}^{2}>0}d\tilde{g}(q)d\tilde{\bar{g}}(p)\left(\frac{\tilde{g}(q)\tilde{\bar{g}}(p)+\tilde{Y}^{2}}{\tilde{Y}}\right)^{\Delta-2}\frac{1}{[(\tilde{g}(q)-\tilde{g}(\zeta))(\tilde{\bar{g}}(p)-\tilde{\bar{g}}(\bar{\zeta}))]^{\Delta}}
\end{align} }
Where in the second line we used the transformation property of conformal primary operator $\mathcal{O}(q,p)\rightarrow\mathcal{O}(g(q),\bar{g}(p))$. In the third line, we defined $\tilde{g} = g -X_{0}$. We can choose $q=-t'+iy'=z$,$p=t'+iy'=-\bar{z}$\footnote{This choice guarantees a gauge choice in bulk spacetime, where the asymptotic limit of that is just Poincare $AdS_{3}$.}, the integration variables $g(q)$ and $\tilde{g}(p)$ are two complex conjugates with overall negative sign. Therefore, we can write $\tilde{g}(q)=re^{i\theta}$ and $\tilde{\bar{g}}(p) = -re^{-i\theta}$. The integration range can also be changed to $\tilde{g}(q)\tilde{\bar{g}}(p)+\tilde{Y}^{2}>0$. Using these polar coordinates we can compute the integral (in a similar way of \cite{Kabat:2011rz}) and the integrand is same as that of $AdS_{3}$ bulk-boundary two point function. Hence the final answer we get is,
\begin{align}\label{bb}
\phi(\tilde{Y},0,0)\mathcal{O}(\zeta,\bar{\zeta})|_{vac} = \left(\frac{\tilde{Y}\sqrt{g'(\zeta)\bar{g}'(\zeta)}}{\tilde{Y}^{2}+(g(\zeta)-X_{0})(\bar{g}(\zeta)-X_{0})}\right)^{2\Delta}
\end{align}
One can also find this vacuum sector of bulk-boundary OPE block from the $AdS_{3}$($y,\omega,\bar{\omega}$) bulk-boundary propagator $G_{b\partial} = \phi(y,0,0)\mathcal{O}(\omega,\bar{\omega}) = \left(\frac{y}{y^{2}+\omega\bar{\omega}}\right)$ \cite{Anand:2017dav}. This is an exact result of symmetry. One can apply a generic operator-valued diffeomorphism of the kind (\ref{diff}) to obtain (\ref{bb}).Using the similar lines of \cite{Anand:2017dav}, one can further compute the $\frac{1}{c}$ corrections of OPE block, which consists of corrections like $T\mathcal{O}$,$TT\mathcal{O}$,$\bar{T}\mathcal{O}$ etc in every order. The zeroth order correction precisely gives the same bulk-boundary two point function $G_{b\partial}$ of $AdS_{3}$.
\subsection{$\frac{1}{c}$ correction of bulk field}\label{TO correction}
 In this section we will derive the $\frac{1}{c}$ correction of that local bulk field in terms of CFT operators. As this field interacts with gravity only, we will expect to be of the form,
 \begin{align}
 \phi &= \phi^{(0)} + \phi^{(1)} + \phi^{(2)} + \dots \nonumber \\
 & = \int K^{(0)} \mathcal{O} + \frac{1}{c} \int K^{(1)}(T)\mathcal{O} + \frac{1}{c^{2}} \int K^{(2)} (TT)\mathcal{O} + \dots
 \end{align}
 Where $K^{(0)}$ is the Kernel of the free $AdS$ field in the vacuum. Here we will recover this and will determine the Kernel of the first correction, $K^{(1)}$. First we show how this can be done in a simplified setting and next we will do it in radial gauge. Let us proceed by considering the simplest example of time slice preserving large spatial diffeomorphism from vacuum $AdS_{3}$ coordinate $(y,x)$ to $(Y,X)$ via:
\begin{align}
x =& g(X) + G(g(X),Y) \quad \ni \quad G(g(X)=X,Y) = 0; \quad G(g(X),Y=0) = 0  \\
y =& YF(g(X),Y) \quad \ni \quad F(g(X)=X,Y) = 1
\end{align}
To compare the geodesic equation on the metric, obtained by this diffeomorphism, we showed that $X_{0} = g(X)+G(g,Y)$, is the bulk point where two geodesics intersect. Therefore we got the following expression for bulk field:
\begin{align}\label{phi non-pert}
\phi(X,Y) = \int dq dp \left(\frac{(g(q)-g(X)-G(g,Y))(g(p)-g(X)-G(g,Y))+Y^{2}F^{2}(g,Y)}{YF\sqrt{g'(q)g'(p)}}\right)^{\Delta-2} \mathcal{O}(p,q)
\end{align}
Where the integration region is bounded by $(g(q)-g(X)-G(g,Y))(g(p)-g(X)-G(g,Y))+Y^{2}F^{2}>0$. The stress tensor for CFT$_{2}$ can be seen as a function of $g(X)$ through the Schwarzian derivative $S(g,X)$.
\begin{align}\label{sch}
S(g,X) = \frac{g'''(X) - \frac{3}{2}(g''(X))^{2}}{(g'(X))^{2}} = \frac{12}{c}T(X)
\end{align}
We need to solve this equation to determine $g(X)$ as a functional of $T(X)$. To do this we first need to write $g(X)$ as a perturbation series in $\frac{1}{c}$. 
\begin{align}\label{g}
g(X) = X + \frac{1}{c}g_{1}(X) + \frac{1}{c^{2}}g_{2}(X) + \dots
\end{align}
Using this perturbation we can also expand $G,F$ as,
\begin{align}\label{G,F}
& G(g,Y) = \sum_{n=1}a_{n}(g(X)-X)^{n}G^{(n)}(X,Y) = \frac{1}{c}a_{1}g_{1}(X)G^{(1)}(X,Y) + \mathcal{O}(\frac{1}{c^{2}}) \nonumber  \\
& F(g,Y) = 1+\sum_{n=1}b_{n}(g(X)-X)^{n}F^{(n)}(X,Y) = 1+ \frac{1}{c}b_{1}g_{1}(X)F^{(1)}(X,Y) + \mathcal{O}(\frac{1}{c^{2}})
\end{align}
Therefore putting (\ref{g}) and (\ref{G,F}) in (\ref{phi non-pert}) we get the kernel upto $\mathcal{O}(\frac{1}{c})$,
\begin{align}
K(p,q,X,Y) = \left(\frac{\psi(q,X,Y)\psi(p,X,Y)+Y^{2}\chi(X,Y)}{Y\left(1+\frac{1}{c}(b_{1}g_{1}(X)F^{(1)}(X,Y)+\frac{g'_{1}(q)+g'_{1}(p)}{2})\right)}\right)^{\Delta-2}
\end{align}
Where
\begin{eqnarray}
&& \psi(q,X,Y) = q+\frac{1}{c}g_{1}(q)-X-\frac{1}{c}g_{1}(X)-\frac{1}{c}a_{1}g_{1}(X)G^{(1)}(X,Y), \nonumber \\
&&\chi(X,Y) = 1+\frac{2b_{1}}{c}g_{1}(X)F^{(1)}(X,Y))
\end{eqnarray}
After some steps of simple algebra we finally get,
\begin{align}
K(p,q,X,Y) = \left(\frac{(q-X)(p-X)+Y^{2}}{Y}\right)^{\Delta-2}\left(1+\frac{1}{c}\mathcal{H}(q,p,X,Y)\right)^{\Delta-2}
\end{align}
Here 
\begin{align}
\mathcal{H} &= \frac{1}{(q-X)(p-X)+Y^{2}}[(p-X)g_{1}(q)+(q-X)g_{1}(p)-(p+q-2X)(a_{1}g_{1}(X)G^{(1)}(X,Y)+g_{1}(X))+ \nonumber \\
& 2Y^{2}b_{1}g_{1}(X)F^{(1)}(X,Y)-((q-X)(p-X)+Y^{2})\left(b_{1}g_{1}(X)F^{(1)}(X,Y)+\frac{g_{1}'(q)+g_{1}'(p)}{2}\right)]
\end{align}
Using the methods discussed in \cite{Anand:2017dav} we can solve (\ref{sch}) and get,
\begin{align}
g_{1}(X) = 6\int_{0}^{X}dX'(X-X')^{2}T(X')
\end{align}
 Therefore we get the perturbative field expansion as,
\begin{align}
\phi(X,Y)= \int \left(\frac{(q-X)(p-X)+Y^{2}}{Y}\right)^{\Delta-2} \mathcal{O}(p,q) + \frac{1}{c}\int   K^{(1)}(T(X'))\mathcal{O}(p,q)
\end{align}
where,
\begin{align}
&K^{(1)}(q,p,X,Y) \nonumber \\
&= (\Delta-2) \left(\frac{(q-X)(p-X)+Y^{2}}{Y}\right)^{\Delta-2}[ \frac{1}{(q-X)(p-X)+Y^{2}} [(p-X)\int^{q}_{0}dX'(q-X')^{2}+\nonumber\\
&(q-X)\int^{p}_{0}dX' (p-X')^{2} -((p+q-2X)(a_{1}G^{(1)}(X,Y)+1)+2Y^{2}b_{1}F^{(1)}(X,Y))\int^{X}_{0}dX' (X-X')^{2}]+ \nonumber \\
&\left(b_{1}F^{(1)}(X,Y)\int^{X}_{0}dX' (X-X')^{2}+\int^{q}_{0}dX'(q-X') + \int^{p}_{0}dX'(p-X')\right)]T(X')
\end{align}
Here, in the both zeroth and $\frac{1}{c}$ term, the integration region is bounded by $(q-X)(p-X)+Y^{2}>0$. But for the case when $\Delta=2$, it will get modified. We will comment on this in the next case.

For the special case, where in the Fefferman-Graham coordinate and in the radial gauge($g_{\mu Y} = 0$)\cite{Banados:1998gg}, the most general solution obeying Brown-Henneaux boundary conditions \cite{Brown:1986nw}, are the special class of diffeomorphism, which takes the following form \cite{Roberts:2012aq},\cite{Anand:2017dav},
\begin{align}
G(g,\bar{g},Y) &= -\frac{2Y^{2}(g'(z))^{2}\bar{g}''(\bar{z})}{4g'(z)\bar{g}'(\bar{z})+Y^{2}g''(z)\bar{g}''(\bar{z})}, \; \bar{G}(g,\bar{g},Y) = -\frac{2Y^{2}(\bar{g}'(\bar{z}))^{2}g''(z)}{4g'(z)\bar{g}'(\bar{z})+Y^{2}g''(z)\bar{g}''(\bar{z})}, \nonumber \\
F(g,\bar{g},Y)& = \frac{4(g'(z)\bar{g}'(\bar{z}))^{\frac{3}{2}}}{4g'(z)\bar{g}'(\bar{z})+Y^{2}g''(z)\bar{g}''(\bar{z})}
\end{align}
In the large $c$ limit the $\frac{1}{c}$ correction of the above functionals are as follow:
\begin{align}\label{radialgauge}
G(g,\bar{g},Y) = -\frac{Y^{2}}{2c}\bar{g_{1}}''(\bar{z}),\; \bar{G}(g,\bar{g},Y) = -\frac{Y^{2}}{2c}g_{1}''(z),\; F(g,\bar{g},Y) = 1+\frac{1}{2c}(g_{1}'(z)+\bar{g_{1}}'(\bar{z}))
\end{align}
Here, we want to compute the $\frac{1}{c}$ correction of the bulk field,
\begin{align}\label{mainexp}
\phi =& \int dq dp \quad \theta\left((g(q)-g(z)-G(g,\bar{g},Y))(\bar{g}(p)-\bar{g}(\bar{z})-\bar{G}(g,\bar{g},Y))+Y^{2}F^{2}(g,Y)\right) \times \nonumber \\
&\left(\frac{(g(q)-g(z)-G(g,\bar{g},Y))(\bar{g}(p)-\bar{g}(\bar{z})-\bar{G}(g,\bar{g},Y))+Y^{2}F^{2}(g,Y)}{YF\sqrt{g'(q)g'(p)}}\right)^{\Delta-2} \mathcal{O}(p,q)
\end{align}
Here for simplicity, we restrict ourselves to holomorphic contribution due to $g(z)$(with $\bar{g}(\bar{z})=0$)only. Using (\ref{radialgauge}) and (\ref{g}) the Kernel $K^{(1)}(q,p,z,\bar{z})$ gives,
\begin{align}\label{k1}
K^{(1)} & = \left(\frac{(q-z+\frac{1}{c}g_{1}(q)-\frac{1}{c}g_{1}(z))(p-\bar{z}+\frac{Y^{2}}{2c}g_{1}''(z))+Y^{2}+\frac{Y^{2}}{c}g_{1}'(z)}{Y\left(1+\frac{1}{2c}(g_{1}'(q)+g_{1}'(z))\right)}\right)^{\Delta-2} \nonumber \\
& = \left[ \frac{\left(\Delta x^{+}\Delta x^{-}+Y^{2}+\frac{1}{c}(\frac{Y^{2}\Delta x^{+}}{2}g_{1}''(z)+\Delta x^{-}(g_{1}(q)-g_{1}(z))+Y^{2}g_{1}'(z))\right)\left(1-\frac{1}{2c}(g_{1}'(z)+g_{1}'(q))\right)}{Y}\right]^{\Delta-2} \nonumber \\
& = \frac{1}{c}K^{(0)}2(h-1)\left[\frac{\frac{Y^{2}\Delta x^{+}}{2}g_{1}''(z)+\Delta x^{-}(g_{1}(q)-g_{1}(z))+Y^{2}g_{1}'(z))}{\Delta x^{+}\Delta x^{-}+Y^{2}} - \frac{1}{2}(g_{1}'(z)+g_{1}'(q))\right] \nonumber \\
& =\frac{1}{c} (h-1)K^{(0)} \left(-\psi^{(2)}(q,z)+\frac{2\Delta x^{+}}{\Delta x^{+}\Delta x^{-}+Y^{2}}\psi^{(3)}(q,z)\right) 
\end{align}
Where $\Delta x^{+} = (q-z)$,$\Delta x^{-} = (p-\bar{z})$. Also we define, $\psi^{(2)}(q,z) = (g_{1}'(q)-g_{1}'(z)-\Delta x^{+}g_{1}''(z))$ and $\psi^{(3)}(q,z) = (g_{1}(q)-g_{1}(z)-\Delta x^{+}g_{1}'(z)-\frac{\Delta x^{+2}}{2}g_{1}''(z))$. As mentioned earlier $K^{(0)}$ is the Kernel of the vacuum $AdS_{3}$. Now, let us look at the integration region, which can be absorbed in a step function, as in (\ref{mainexp}). This step function, inside the integral, could also get correction in such $\frac{1}{c}$ expansion as follows:
\begin{align}\label{thetaecpansion}
&\theta\left((g(q)-g(z)-G(g,\bar{g},Y))(\bar{g}(p)-\bar{g}(\bar{z})-\bar{G}(g,\bar{g},Y))+Y^{2}F^{2}(g,Y)\right) \nonumber \\
&=\theta\left((q-z)(p-\bar{z})+Y^{2}\right) + \frac{1}{c}\left(\frac{Y^{2}\Delta x^{+}}{2}g_{1}''(z)+\Delta x^{-}(g_{1}(q)-g_{1}(z))+Y^{2}g_{1}'(z)\right)\delta\left((q-z)(p-\bar{z})+Y^{2}\right)
\end{align}
Using this expansion we can finally get the following expression for the $\frac{1}{c}$ correction of bulk field, in terms of (\ref{k1}), as,
\begin{align}
\phi^{(1)}(X,Y) = \int \theta\left((q-z)(p-\bar{z})+Y^{2}\right) K^{(1)}(q,p,X,Y) \mathcal{O}(p,q)
\end{align}
 This is the expression we can get from the diffeomorphism in the radial gauge, as described in \cite{Guica:2016pid}. Next it is straightforward to express $\psi^{(2)}$ and $\psi^{(3)}$ in terms of $T_{++}(x')$ and to check that it satisfies the linearised equation of motion,
\begin{align}
(\nabla^{2} - m^{2})\phi^{(1)} = -\frac{24z^{4}}{c}T_{++}\delta_{\bar{z}}^{2}\phi^{(0)}
\end{align}
But the expression is only valid when $\Delta \neq 2$. For the $\Delta=2$ case, we get contribution from the $\frac{1}{c}$ correction of step function. In that case, the full expression for $\phi$ upto $\frac{1}{c}$ term, is the following,
\begin{align}
\phi^{(\Delta=2)}&=\int dq dp \quad \theta\left((q-z)(p-\bar{z})+Y^{2}\right) \mathcal{O}(p,q)+ \nonumber \\
& \frac{1}{c}\int dq dp \quad \delta\left((q-z)(p-\bar{z})+Y^{2}\right)\left(\frac{Y^{2}\Delta x^{+}}{2}g_{1}''(z)+\Delta x^{-}(g_{1}(q)-g_{1}(z))+Y^{2}g_{1}'(z)\right) \mathcal{O}(p,q)
\end{align}
In similar fashion, one might extend this analysis to compute all the higher order correction in $\mathcal{O}(\frac{1}{c^{2}})$.

\vspace*{1ex}
\noindent{\bf Acknowledgment:} 
We would like to thank Prof Gilad Lifschytz for reading our manuscript and for illuminating discussions and pointing out concrete corrections to our results as well as general suggestions about the  paper. This research was supported in part by the International Centre for Theoretical Sciences (ICTS) during a visit by one of us (BE), to  participate in the program - AdS/CFT at 20 and Beyond (Code: ICTS/adscft20/2018/05). The work of SD was supported by a senior research fellowship(SRF) from CSIR.

\end{document}